\documentclass{article}
\usepackage[dvips]{graphicx}
\begin{document}

\begin{center}
\centerline{\large \bf Should physicists begin experimental study}
\centerline{\large \bf of the God's physical nature?} 
\end{center}

\vspace{3 pt}
\centerline{\sl V.A. Kuz'menko\footnote{Electronic 
address: kuzmenko@triniti.ru}}

\vspace{5 pt}

\centerline{\small \it Troitsk Institute for Innovation and Fusion 
Research,}
\centerline{\small \it Troitsk, Moscow region, 142190, Russian 
Federation.}

\vspace{5 pt}

\begin{abstract}

Inequality of forward and reversed processes in quantum physics means an 
existence of a memory of quantum system about the initial state. Importance 
of its experimental study for correct interpretation of quantum mechanics 
and understanding of a physical base of a consciousness is discussed.

\vspace{12 pt}
\end{abstract}

\vspace{8 pt}

Physics and religion practically do not intersect. Modern religion supposes 
an existence in nature some delocalized Superior Mind. Unfortunately, 
there is absent a matter for physical study here: it is unclear what its 
physical base is. However, the task to find in quantum physics a base for 
consciousness and mind is important for physicists and biologists [1]. 
The main problem here is that the physicists till now do not sufficiently 
understand a fundament of quantum physics itself. Quantum mechanics is a 
mathematical model which gives excellent description of the result of 
quantum process, but it nothing says about the process itself, its nature. 
Due to this reason Einstein believed the quantum mechanics as an incomplete 
theory. Hot debates about the interpretation of quantum mechanics last till 
now with unflagging power [2 - 5]. The number of interpretations of quantum 
mechanics exists today: the Copenhagen interpretation [6], the theory of 
hidden parameters of Bohm [7], the many-worlds interpretation [8], the 
consistent histories interpretation [9] and others [10]. The existence of 
so numerous competitive interpretations indicates about the condition of 
deep and permanent stagnation in quantum physics (in spite of superficial 
sings of development). It is unclear in this situation which theory is 
right and may be used as a base for understanding and the study of a nature 
of consciousness and mind.

We shall come to this problem from other side. In the field of high energy 
physics the scientists long ago recognized the fact of CP and T invariance 
violation in physics [11]. However, quite opposite situation exists in 
quantum physics. Many years the common opinion here is that "... a remarkable 
fundamental fact of nature: all known laws of physics are invariant under 
time reversal" [12]. Overwhelming majority of physicists follows this 
paradigm till now. But this is a great error [13] and, perhaps, the main 
obstacle today on the way of progress in quantum physics.

In contrast to this paradigm we have several direct and great numbers of 
indirect experimental proofs of inequality of forward and reversed processes 
in quantum physics [14]. Although the integral cross-sections of such 
transitions should be equal, its differential cross-sections can differ in 
several orders of magnitude. The reversed into the initial state transition 
has much more high differential cross-section, but and much more sharp 
dependence from different physical parameters (Fig.1). As a fact, such 
difference is a physical base of most phenomena in nonlinear optics. Here 
the situation is quite similar to those in whole quantum physics: we have 
the number of mathematical models which usually give good descriptions of 
phenomena, but with its indistinct physical explanations. 
	
Inequality of forward and reversed processes supposes an existence of some 
memory of quantum system about its initial state. Without such memory the 
quantum system cannot define next process as forward or reversed. However, 
it is unclear now what physical bearer of this memory is and where it is 
located or delocalized. We have some reason to believe that this memory may 
be nonlocal in a substantial degree. Quite unusual argument outside of 
science testifies to such possibility. So-called person with extrasensory 
perception can extract information about the past events from the 
environment. They do it better on the place of the event. But this is 
possible also (with less efficiency) at great distance away from that place, 
which points to possible substantial nonlocality of such kind of the 
environment's memory. 

The Bohm's theory is a variant of interpretation of quantum mechanics [7]. 
It supposes an existence of some hidden variables, which need for complete 
description of quantum system and some exotic nonlocal quantum potential. 
Today this theory seems to be not very popular among the physicists [10], 
probably, because it is unclear what is a nature of supposed hidden 
variables and nonlocal quantum potential. However, we can see now that 
differential cross-sections in Fig.1 may be a beautiful example of hidden 
parameters for processes of absorption and scattering of photons by atoms 
and molecules. From other side the memory of quantum system about its 
initial state looks like as a direct equivalent of the nonlocal quantum 
potential. So, the recognizing of inequality of forward and reversed 
processes in quantum physics quite apparently leads to conclusion that the 
Bohm's theory is the most correct interpretation of quantum mechanics. 
And nonlocal quantum potential (memory) is a good physical base for 
explanation of a nature of consciousness and mind in the universe.

Unfortunately, we cannot measure such quantum potential (memory) today. 
However, the extrasensory individuals can do it. Although, they do not 
understand how they do it. But we can measure the differential 
cross-sections of quantum transitions. It, possibly, will lead in a future 
to a better understanding of a nature of quantum memory and to technical 
capability to extract such memory from the environment.

The most convenient object today for experimental study of differential 
cross-sections of quantum transitions is the so-called Bloch oscillations 
of cold atoms in vertical optical lattice [15, 16]. Here the cold atoms 
under action of gravity freely fall down in vacuum. In certain point the 
specific scattering of photons takes place: one upward photon is absorbed 
and one downward photon is emitted. As a result, the recoil momentum 
returns the atoms exactly into the initial point (state). This is the most 
clean example of fully reversed quantum transition today. This transition 
has highest possible differential cross-section. All other possible 
processes will be partially reversed or forward and will have lesser 
differential cross-sections (Fig.1). The dependences of the differential 
cross-sections from physical parameters (frequency and phase of laser 
radiation, its direction, the position of the atom in space) nobody 
experimentally studied till now [17]. This is not very difficult task. 
However, the main problem here is that before such experiments our 
physicists must reject their paradigm: "... all known laws of physics are 
invariant under time reversal".

\begin{figure}
\newpage
\begin{center}
\includegraphics[scale=1.0]{Figure1.ps}
\end{center}
\end{figure}

\end{document}